\documentclass[a4paper,10pt]{article}

\usepackage{graphicx}
\usepackage{amsmath,amssymb,booktabs}
\usepackage{epsf,epsfig}
\usepackage{psfrag}
\usepackage{cite}
\usepackage{mathrsfs}
\usepackage{bbm}
\usepackage[caption=false]{subfig}
\usepackage{dcolumn}
\usepackage[export]{adjustbox}
\allowdisplaybreaks

\def\be{\begin{equation}}
\def\ee{\end{equation}}
\def\beq{\begin{eqnarray}}
\def\eeq{\end{eqnarray}}

\def\DB0{\partial B_0}

\def\Cl2{\mbox{Cl}_2}

\usepackage{color}

\definecolor{Brown}{rgb}{0.5,0.25,0}

\begin{document}


\begin{flushright}
\today
\end{flushright}
\vskip 1.0cm

\begin{center}
\Large{\bf\boldmath
Diagrammatic analysis of hidden- and open-charm tetraquark production in $B$ decays
\unboldmath}

\normalsize
\vskip 0.8cm

{\sc Qin~Qin\footnote{qqin@hust.edu.cn}}
\vskip 0.2cm
{\it School of physics, Huazhong University of Science and Technology, Wuhan 430074, China}\\

\vskip 0.5cm

{\sc Jing-Liang Qiu and}
{\sc Fu-Sheng Yu\footnote{yufsh@lzu.edu.cn, corresponding author}}\\

\vskip 0.2cm

{\it School of Nuclear Science and Technology, Lanzhou University, Lanzhou 730000, China}\\
{\it and Lanzhou Center for Theoretical Physics, and Frontiers Science Center for Rare Isotopes, Lanzhou University, Lanzhou 730000, China}\\
{\it and Center for High Energy Physics, Peking University, Beijing 100871, China}\\

\vskip 0.8cm

\end{center}


\begin{abstract}
Discoveries of tetraquarks not only enrich the hadronic spectrum but also provide more platforms to understand quantum chromodynamics. We study the production processes of hidden-charm and open-charm tetraquarks in $B$ decays by analyzing their topological amplitudes. Relations between different channels are found, which confront tests by experiments to probe the nature of the tetraquarks. Furthermore, promising channels to find more tetraquarks are proposed. 
\end{abstract}


\section{Introduction}
\label{sec:intro}

Since the discovery of $X(3872)$~\cite{Belle:2003nnu}, more and more exotic hadronic states have been observed in experiments (see~\cite{Chen:2016qju,Chen:2016spr,Liu:2019zoy,Guo:2017jvc,Chen:2022asf} for reviews) and also are expected to found in future~\cite{Yang:2022nps,Qin:2021zqx,Ali:2018xfq,Ali:2018ifm}. 
Together with mesons and baryons, they form a much more colorful and comprehensive "periodic table of hadrons", which provide dramatic platforms for deciphering the quantum chromodynamics. 
Many explanations for them such as multiquark states~\cite{Chiu:2006hd,Bigi:2005fr,Maiani:2005pe}, molecular states~\cite{Voloshin:1976ap,Bander:1975fb,DeRujula:1976zlg}, hybrid mesons~\cite{Horn:1977rq,Close:1994zq,McNeile:2002az} and missing excited hadron states~\cite{Barnes:2003vb,Eichten:2004uh} are proposed. Later, $Z^\pm(4430)$ was observed~\cite{Belle:2007hrb} and confirmed~\cite{LHCb:2014zfx} in the $B$ meson decay channel $B^0 \to K^\mp Z^\pm(4430), \ Z^\pm(4430)\to \pi^\pm \psi'$, and its partner particles $Z_c^+(4200)$~\cite{Belle:2014nuw}, $Z_c^+(4050)$ and $Z_c^+(4250)$~\cite{Belle:2008qeq} were also reported, which definitely have structures containing four quarks, beyond the explanation of hybrid mesons or an excited charmonia.\footnote{The discovery of $T_{cc}^+$~\cite{Qin:2020zlg,LHCb:2021vvq,LHCb:2021auc} also strengthens this conclusion.}
More of such tetraquarks were discovered in $B$ meson decays, such as $T^\theta_{\psi s1}(4000)^+$ and $T_{\psi s1}(4220)^+$~\cite{LHCb:2021uow}, and $T_{cs0,1}[cs\bar{u}\bar{d}]$~\cite{LHCb:2020pxc}. Very recently, two new $D_s^+\pi^\pm$ resonances $T^a_{c\bar{s}0}(2900)^{++}[c\bar{s}{u}\bar{d}]$ and $T^a_{c\bar{s}0}(2900)^0[c\bar{s}\bar{u}{d}]$\footnote{See~\cite{Gershon:2022xnn} for a name convention for exotic hadrons.} were found in~\cite{LHCb:2022xob,LHCb:2022bkt}
\begin{align}
B^+ \to D^-D_s^+\pi^+ \;, \qquad 
B^0 \to \bar{D}^0 D_s^+\pi^- \; ,
\end{align}
respectively. Such tetraquark candidates were studied by previous theoretical works~\cite{He:2020jna,Lu:2020qmp,Burns:2020xne,Agaev:2021knl,Agaev:2021jsz,Azizi:2021aib,Yue:2022mnf}.

It can be found that the $B$ decays are a good place for the tetraquark production\cite{An:2022vtg,Chen:2020eyu,Yu:2017pmn}. In this work, we attempt to probe the mechanism of such tetraquark production processes via $B$ meson decays, employing the framework of the topological amplitude. 
We do not consider the states containing a $q\bar{q}$ ($q=u,d,s$) pair, since they are not easy to be identified from ordinary meson states, and also suffer large backgrounds from ordinary states.

The dynamics of the tetraquark productions in $B$ decay are so complicated that it is nearly impossible to perform the first principle calculation. In such a case, the diagrammatic approach can nevertheless teach us essential information about the physical processes. The diagrammatic approach has been applied in $B$ decays~\cite{Zhou:2015jba,Zhou:2016jkv,Huber:2021cgk,Zhou:2021yys} and $D$ decays~\cite{Bhattacharya:2009ps,Cheng:2010ry,Li:2012cfa,Qin:2013tje,Jiang:2017zwr,Qin:2021tve,Yu:2017oky,Wang:2017ksn}. It has also been proved to be able to recover the symmetry protected relations between different decay amplitudes~\cite{He:2018php,Wang:2020gmn}. 
We apply it to the $B$ decay channels into a tetraquark and a meson. Such an analysis can not give explicit values of decay amplitudes, but can build relations between different decay modes, which will be useful in at least two aspects. Firstly, with the information of observed channels, we can predict which tetraquarks have great potential to be discovered with the current database and which channels should be used for the search. Secondly, the compact tetraquark structure is assumed to obtain some relations between different channels, so the experimental tests of these relations can examine validity of the assumption. The production of heavy-flavor tetraquarks in $B$ decays and their decays have also been studied based on the flavor SU(3) symmetry analysis~\cite{He:2016xvd,Xing:2019hjg,He:2020jna,Huang:2022zsy}.

The hidden-charm and open-charm tetraquark production processes in $B$ decays are studied. For the hidden-charm case, the $b\to c\bar c s$ processes are considered, which have larger Cabbibo-Kobayashi-Maskawa (CKM) matrix elements than other production processes. For the open-charm case, both the $b\to c\bar c s$ and $b\to c\bar u d$ processes are considered, because both of them are CKM favored. The topological amplitudes of all the processes are given, and it is found that they can be classified into several groups, in which the processes share the same topological amplitude. The decay processes in the same group are expected to have similar branching ratios, and if one of them have been discovered experimentally, the others should also have a large potential to be observed. 

The rest of the paper is organized as follows. In section~\ref{sec:topo}, the hidden-charm and open-charm tetraquark production processes in $B$ meson decays are analyzed by using their topological amplitudes. We conclude by section~\ref{sec:con}.

\section{Topological amplitude analysis}\label{sec:topo}

\begin{figure}[htbp]
\centering
\includegraphics[width=0.25\linewidth]{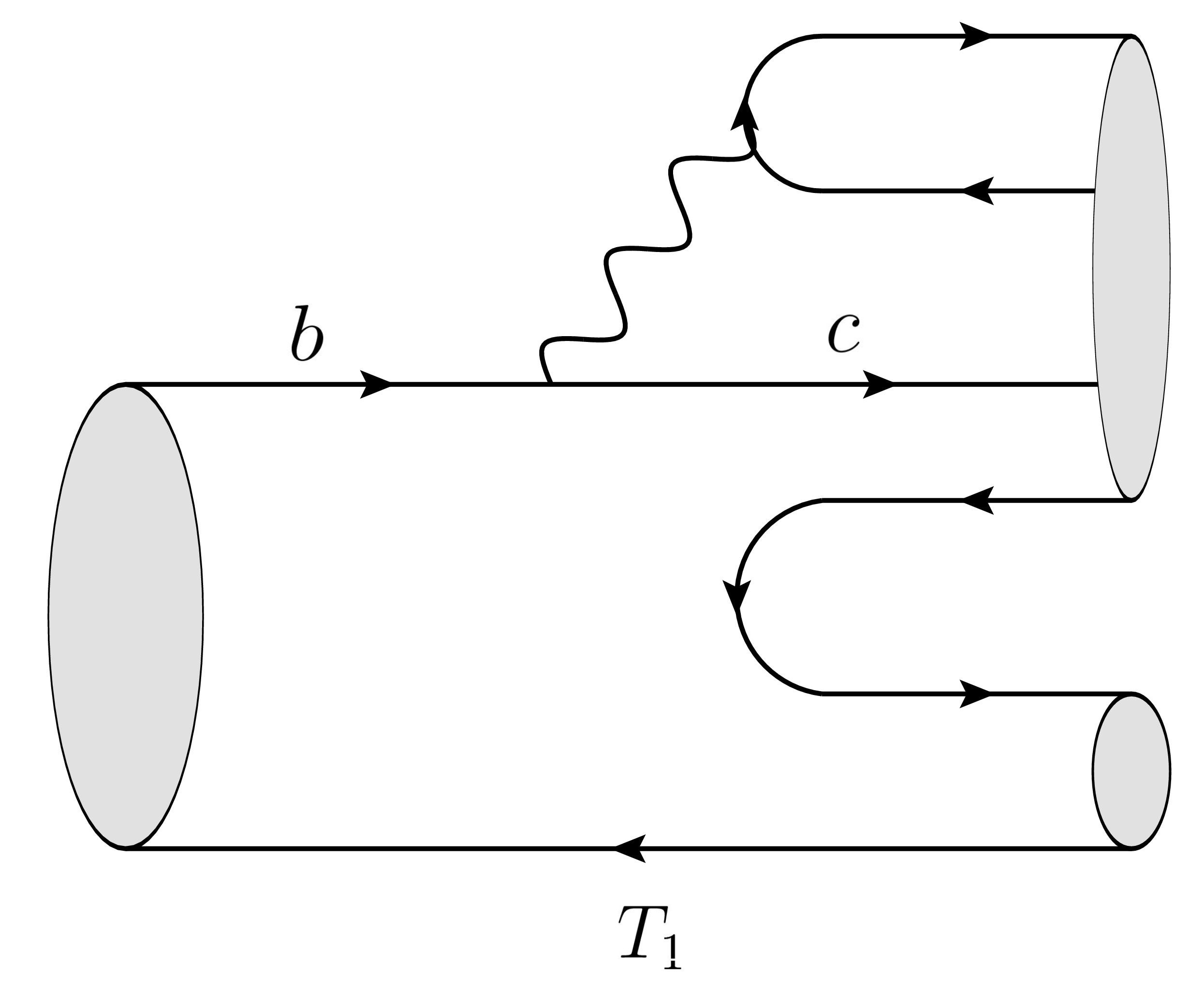}
\includegraphics[width=0.3\linewidth]{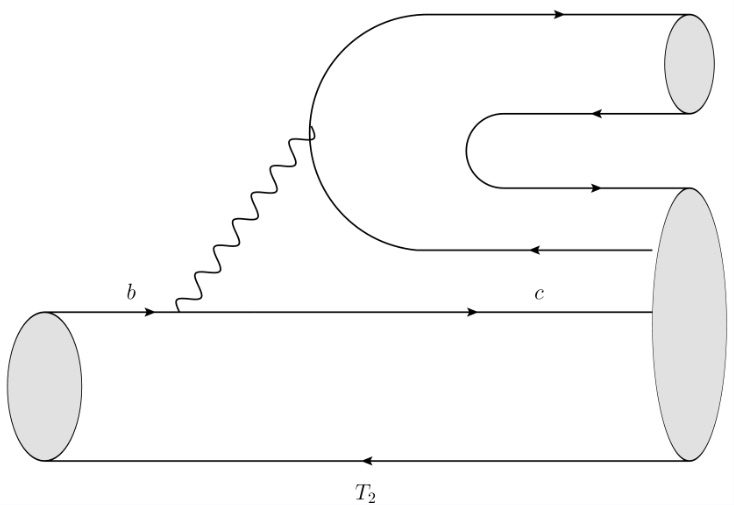}
\includegraphics[width=0.3\linewidth]{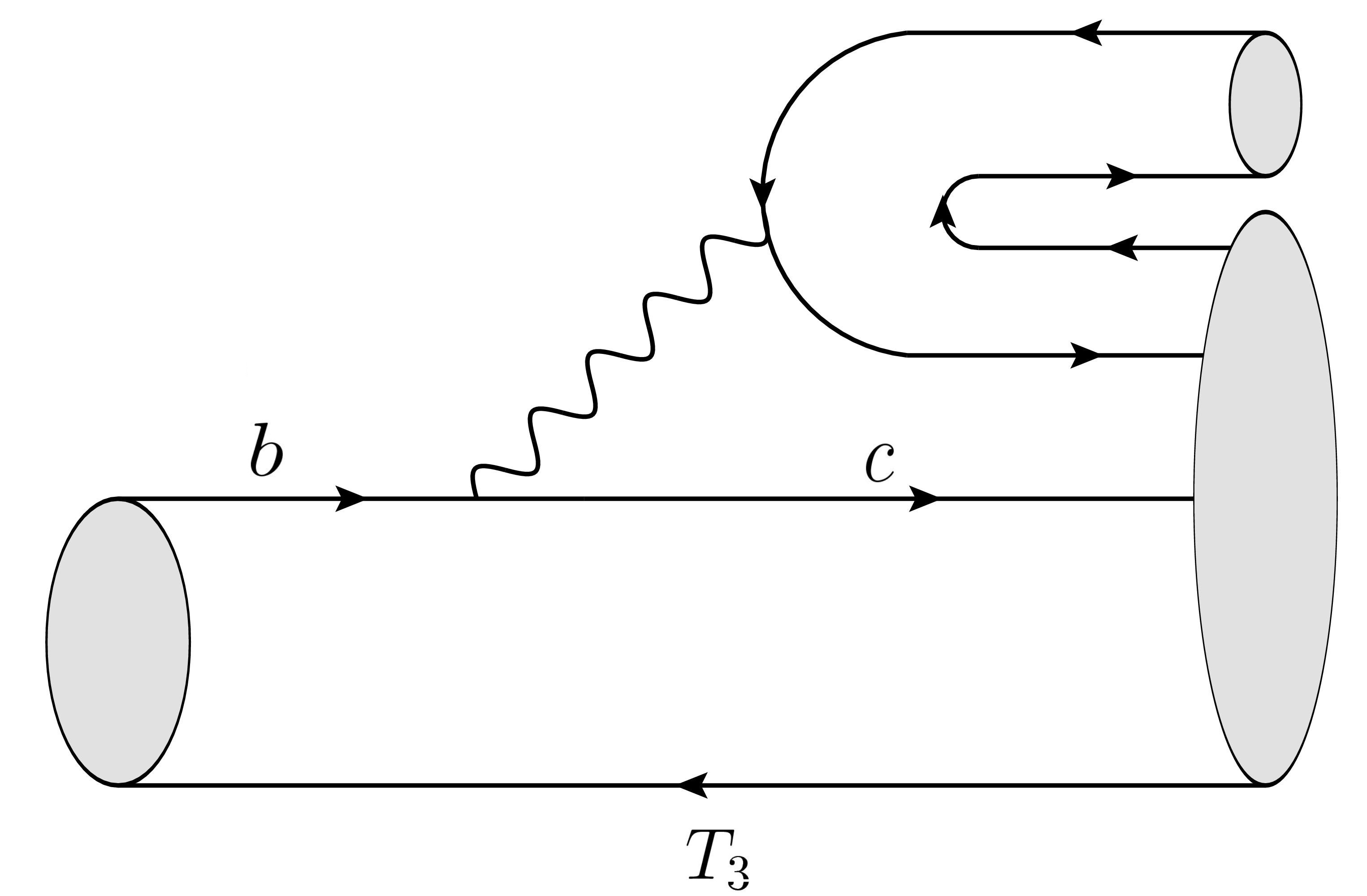}
\includegraphics[width=0.3\linewidth]{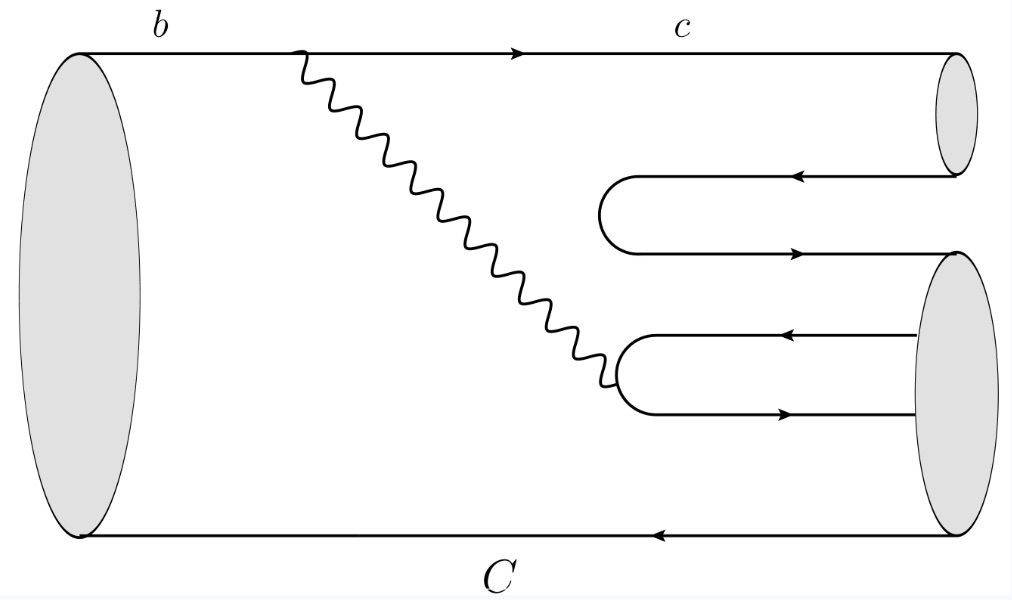}
\includegraphics[width=0.3\linewidth]{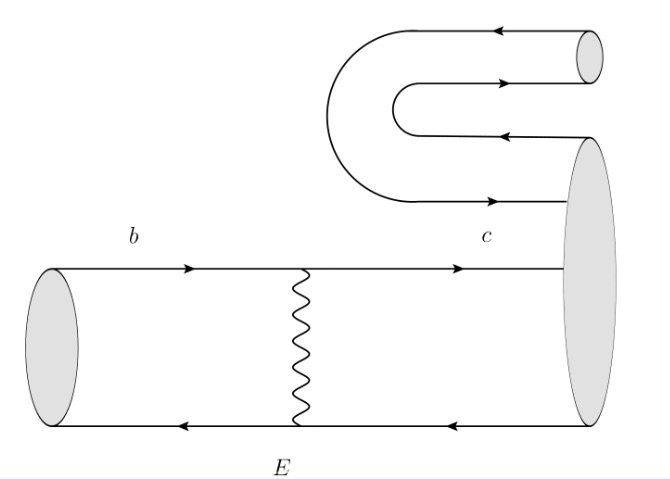}
\caption{Topological diagrams that contribute to the $B$ meson decays into a tetraquark and a meson. The wavy lines indicate the $W$ boson exchanged in the four-quark vertices. All possible gluon exchanges between the quark lines are in principle included but omitted in the diagrams.}
\label{fig:ta}
\end{figure}

The tetraquarks are produced in $B$ decays via weak interactions, so typically a four-fermion weak-interaction vertex is involved in the effective theory of $b$ decays. For simplicity, we only consider the most frequent processes induced by tree-level quark transition with large CKM matrix elements, namely, the $b\to c\bar{c}s$ and $b\to c \bar{u}d$ processes. The most promising channels at experiments should be among them, since the involved CKM matrix elements $V_{cb}$ is the largest $b$-decay one and $V_{ud,cs}\sim 1$. 
Based on these weak transitions, five kinds of topological amplitudes contribute to the processes of the $B$ meson (including $B^-,B^0,B_s^0$) decaying into a tetraquark (including the hidden-charm tetraquarks $c\bar{c}q_1\bar{q}_2$ and the open-charm tetraquarks $cq_1\bar{q}_2\bar{q}_3$ with the light quarks $q_{i} =u,d,s$) and a meson (including charmed mesons and light mesons), as shown in Figure~\ref{fig:ta}, labelled as $T_{1,2,3}$, $C$ and $E$. The quark lines disconnected to the weak-interaction vertices are produced by pair from the vacuum or transited from the spectator quark in the initial-state $B$ meson. Note that the tetraquarks containing a $\bar{q}q$ pair are not considered, so the diagrams with a vacuum $q\bar{q}$ pair entering a tetraquark are not included.

If two decay channels share the identical topological amplitude, they are expected to have analogous magnitudes of decay amplitudes, even if the final-state particles have different spins or belong to different flavor SU(3) representations, which was observed in mesonic $B$ meson decays~\cite{Zhou:2015jba,Zhou:2016jkv}. Therefore, these decay channels also have anomalous branching ratios. There might be exceptions owing to some special symmetry rules such as isospin cancellation and helicity suppression. Next, we will analyze the topological amplitudes of the $B$ decays into a hidden-charm tetraquark or an open-charm tetraquark.

\subsection{Hidden-charm tetraquarks}

\begin{table}[!h]
\caption{Topological amplitudes of the $B$ meson decaying into one hidden-charm open-flavor tetraquark and a meson.}\label{tb:hiddenc}
\begin{tabular}{lcllccccc}\hline\hline
Modes & Topological amplitudes & Experimental processes
 & Experimental processes \\\hline
${B^-\to T_{c\bar cs\bar u}\phi }$ &$T_2$ $V_{cb}V_{cs}^*$ & $B^-\to D^{(*)0}D_s^{(*)-}\phi $ & $B^-\to J/\psi K^-\phi $~\cite{LHCb:2021uow}  \\

$\overline B^0\to T_{c\bar cs\bar d}\phi $ &$T_2$ $V_{cb}V_{cs}^*$ & $\overline B^0\to D^{(*)+}D_s^{(*)-}\phi $ & $\overline B^0\to J/\psi \overline K^{(*)0}\phi $  \\

$\overline B^0\to T_{c\bar cu\bar d}K^-$ &$T_2$ $V_{cb}V_{cs}^*$ & $\overline B^0\to D^{(*)+}\overline D^{(*)0}K^-$ & $\overline B^0\to J/\psi \pi^+K^-$~\cite{Belle:2014nuw}  \\

$B^-\to T_{c\bar cd\bar u}\overline K^{(*)0}$ &$T_2$ $V_{cb}V_{cs}^*$ & $B^-\to D^{(*)0}D^{(*)-}\overline K^{(*)0}$ & $B^-\to  J/\psi \pi^-\overline K^{(*)0}$  \\

$\overline B_s^0\to T_{c\bar cd\bar s}\overline K^{(*)0}$ & $(T_2+E)$ $V_{cb}V_{cs}^*$ & $\overline B_s^0 \to D_s^{(*)+}D^{(*)-}\overline K^{(*)0}$ & $\overline B_s^0\to J/\psi K^0\overline K^{(*)0}$  \\

$\overline B_s^0\to T_{c\bar cu\bar s}K^-$ & $(T_2+E)$ $V_{cb}V_{cs}^*$ & $\overline B_s^0 \to D_s^{(*)+}\overline D^{(*)0}K^-$ & $\overline B_s^0\to J/\psi K^+K^-$  \\

$\overline B_s^0\to T_{c\bar cu\bar d}\pi^-$ &$E$ $V_{cb}V_{cs}^*$ & $\overline B_s^0\to D^{(*)+}\overline D^{(*)0}\pi^-$ & $ \overline B_s^0\to J/\psi \pi^+\pi^-$  \\
$\overline B_s^0\to T_{c\bar cd\bar u}\pi^+$ &$E$ $V_{cb}V_{cs}^*$
 
   & $\overline B_s^0\to D^{(*)0}D^{(*)-}\pi^+$ & $\overline B_s^0\to 
 
  J/\psi \pi^-\pi^+$  \\
$\overline B^0\to T_{c\bar cs\bar u}\pi^+$ &$T_1$ $V_{cb}V_{cs}^*$
 
   & $\overline B^0\to D^{(*)0}D_s^{(*)-}\pi^+$ & $\overline B^0\to 
 
  J/\psi K^-\pi^+$  \\
$B^-\to T_{c\bar cs\bar d}\pi^-$ &$T_1$ $V_{cb}V_{cs}^*$ & $B^-\to D^{(*)+}D_s^{(*)-}\pi^-$ & $B^-\to J/\psi \overline K^0\pi^-$  \\

$B^-\to T_{c\bar cs\bar u}\pi^0$ &${1\over\sqrt{2}}T_1$ $V_{cb}V_{cs}^*$ & $B^-\to D^{(*)0}D_s^{(*)-}\pi^0$ & $B^-\to J/\psi K^-\pi^0$  \\

$\overline B^0\to T_{c\bar cs\bar d}\pi^0$ &${1\over\sqrt{2}}T_1$ $V_{cb}V_{cs}^*$ & $\overline B^0\to D^{(*)+}D_s^{(*)-}\pi^0$ & $\overline B^0\to J/\psi \overline K^0\pi^0$  \\

$\overline B_s^0\to T_{c\bar cs\bar d}K^{(*)0}$ &
 
 $(T_1+E)$ $V_{cb}V_{cs}^*$ & $\overline B_s^0
 
  \to D^{(*)+}D_s^{(*)-}K^{(*)0}$ & $\overline B_s^0\to 
 
  J/\psi \overline K^0K^{(*)0}$  \\
$\overline B_s^0\to T_{c\bar cs\bar u}K^+$ & $(T_1+E)$ $V_{cb}V_{cs}^*$ & $\overline B_s^0
 
  \to D^{(*)0}D_s^{(*)-}K^+$ & $\overline B_s^0\to J/\psi K^-K^+$  \\
\hline\hline
\end{tabular} \end{table}

In this section, we focus on the $B$ meson decays into a hidden-charm tetraquark and a meson, $\bar{B}\to T_{c\bar{c}q_1\bar{q}_2} M$. The meson $M$ must be a light meson because at most two $c(\bar{c})$ quarks are generated in the weak-interaction vertices and the heavy $c\bar{c}$ pair production from vacuum is highly suppressed and neglected here. All the decay channels in this category are listed in Table~\ref{tb:hiddenc}, with also the corresponding topological amplitudes and possible experimental processes to search for these channels. The spins or parities of the tetraquarks are not specified, and they may belong to the flavor SU(3) octet or decuplet. The tetraquarks $T_{c\bar{c}q_1\bar{q}_2}$ typically decay strongly into two mesons composed by $c\bar{c}$ and $q_1\bar{q}_2$, or $q_1\bar{c}$ and $c\bar{q}_2$. The light meson could be a ground state pseudoscalar meson or a vector meson. To avoid the troublesome quark flavor mixing in $\eta$ and $\eta'$, we do not consider these two mesons. For the $s\bar{s}$ component, we only consider the $\phi$ meson, and for $u\bar{u}$ and $d\bar{d}$, we consider both $\pi^0$ and $\rho^0$, the latter having a much higher detection efficiency at the LHC. For conciseness, we only display $\pi^0$ while hiding $\rho^0$ in Table~\ref{tb:hiddenc}, but it should be understood in the way that everywhere the $\pi^0$ appears the $\rho^0$ is an alternative option. 

It can be observed from Table~\ref{tb:hiddenc} that all the decays channel can be classified into 5 different categories according to their topological amplitudes $T_2$, $T_2+E$, $E$, $T_1$ and $T_1+E$. The channels in each category have similar branching ratios. Furthermore, because the weak-exchange amplitude $E$ is expected to be small compared to others, the branching ratios of the $T_2$ and $T_2+E$ channels should be at the same order of magnitude, and it also applies to the $T_1$ and $T_1+E$ channels. Besides the topological amplitude argument, some channels are related by the flavor symmetries, including 
\begin{align}
\text{isospin symmetry: } \ &A(B^-\to T_{c\bar cs\bar u}\phi)  = A(\overline B^0\to T_{c\bar cs\bar d}\phi) \; , \\
&A(\overline B^0\to T_{c\bar cu\bar d}K^{(*)-}) = A(B^-\to T_{c\bar cd\bar u}\overline K^{(*)0}) \; ,\\
&A(\overline B_s^0\to T_{c\bar cd\bar s}\overline K^{(*)0}) = A(\overline B_s^0\to T_{c\bar cu\bar s}K^{(*)-})\; ,\\
&A(\overline B_s^0\to T_{c\bar cu\bar d}\pi^-) = A(\overline B_s^0\to T_{c\bar cd\bar u}\pi^+)\; ,\\
&A(\overline B^0\to T_{c\bar cs\bar u}\pi^+) = A(B^-\to T_{c\bar cs\bar d}\pi^-) \; ,\\
&A(B^-\to T_{c\bar cs\bar u}\pi^0) = A( \overline B^0\to T_{c\bar cs\bar d}\pi^0) \; , \\
&A(\overline B_s^0\to T_{c\bar cs\bar d}K^{(*)0}) = A( \overline B_s^0\to T_{c\bar cs\bar u}K^{(*)+}) \; .
\end{align}
The above relations can be tested at future experiments. Because only the $b\to c\bar{c}s$ processes are taken into account, there are no processes related by the $U$-spin or $V$-spin symmetry. 

Resonances of two types of four-quark constituents $T_{c\bar cu\bar d}$ and $T_{c\bar cs\bar u}$ have been observed experimentally in the channels 
$\overline B^0\to T_{c\bar cu\bar d}K^- \to J/\psi \pi^+K^-$~\cite{Belle:2014nuw} and ${B^-\to T_{c\bar cs\bar u}\phi } \to J/\psi K^-\phi$~\cite{LHCb:2021uow}, respectively. It should be noted that the topological amplitude of these two channels are both $T_2$, so it is expected that $T_2$ is large and the $T_2$ and $T_2+E$ channels should be searched with priority. We list the most promising ones as follows,
\begin{align}
&\overline B^0\to T_{c\bar cs\bar d}\phi \to J/\psi \overline K^{(*)0}\phi \; , \\
&B^-\to T_{c\bar cd\bar u}\overline K^{(*)0} \to  J/\psi \pi^-\overline K^{(*)0}  \; , \\
&\overline B_s^0\to T_{c\bar cd\bar s}\overline K^{(*)0} \to J/\psi K^0\overline K^{(*)0} \; , \\
&\overline B_s^0\to T_{c\bar cu\bar s}K^- \to J/\psi K^+K^- \; .
\end{align}

\subsection{Open-charm tetraquarks}

In this section, we study the $B$ meson decays into a open-charm tetraquark and a meson, $\bar{B}\to T_{\bar c\bar{q}_1q_2q_3} M$. 
The meson $M$ can be a charmed meson $D^{(*)0,+}$ or $D_s^{(*)+}$ if the involved quark transition is $b\to c\bar{c}s$, and it can also 
be a light meson as in the hidden-charm tetraquark case if the quark transition is $b\to c\bar{u}d$. 
The open-charm tetraquarks $T_{\bar c \bar{q}_1 q_2 q_3}$ are constituted by one anti-charm quark, one light antiquark and two quarks. 
They are in the following irreducible representations of the flavor SU(3) group, 
\begin{equation}
{\bf 3} \otimes {\bf 3} \otimes {\bf \bar{3}} = {\bf 15} \oplus {\bf \bar{6}}  \oplus {\bf 3}  \oplus {\bf 3} \; .
\end{equation}
Because the $\bar{q}q$ states are omitted, the {\bf 3} representations are excluded. The {\bf 15} representation is 
partially symmetric containing {\it e.g.} the states $uu\bar{s}$ and $\{ud\}\bar{s}$, and the ${\bf \bar{6}}$ representation is 
partially antisymmetric containing  {\it e.g.} the state $[ud]\bar{s}$. Therefore, the tetraquarks with two identical 
light flavors such as $T_{\bar{c}\bar s uu}$ must belong to the {\bf 15} representation, while the tetraquarks with 
three different light flavors such as $T_{\bar{c}\bar s ud}$ can belong to either the {\bf 15} representation or the ${\bf \bar{6}}$ representation. 
All the CKM favored decays channels are listed in Table~\ref{tb:openc}, with the topological amplitudes and possible experimental processes to search for them. 

\begin{table}[!h]
\caption{Topological amplitudes of the $B$ meson decaying into one open-charm tetraquark and a meson.}\label{tb:openc}
\begin{tabular}{lcllccccc}\hline\hline
Modes & Topological amplitudes & Experimental\  processes & Experimental processes \\\hline
$B^-\to T_{\bar c\bar u ds}D^{(*)+}$ &$C$ $V_{cb}V_{cs}^*$ & $B^- \to D_s^{(*)-}\pi^-D^{(*)+}$~\cite{LHCb:2022xob,LHCb:2022bkt} & $B^-\to D^{(*)-}K^-D^{(*)+}$ \\

$\overline B^0\to T_{\bar c\bar d u s}D^{(*)0}$ &$C$ $V_{cb}V_{cs}^*$ & $\overline B^0\to D_s^{(*)-}\pi^+D^{(*)0}$~\cite{LHCb:2022xob,LHCb:2022bkt} & $\overline B^0\to \overline D^{(*)0}\overline K^0D^{(*)0}$  \\

$B^-\to T_{\bar c\bar u s s}D_s^{(*)+}$ &$\sqrt{2} C$ $V_{cb}V_{cs}^*$ & $ B^-\to D_s^{(*)-}K^-D_s^{(*)+}$  \\

$\overline B^0\to T_{\bar c\bar d s s}D_s^{(*)+}$ & $ \sqrt{2} C$ $V_{cb}V_{cs}^*$ & $\overline B^0\to D_s^{(*)-}\overline K^0 D_s^{(*)+}$  \\

$B^-\to T_{cs\bar u\bar d}D^{(*)-}$ &$T_3$ $V_{cb}V_{cs}^*$ & $B^-  \to D^{(*)+}K^-D^{(*)-}$~\cite{LHCb:2020pxc} & $B^-\to D^{(*)0}\overline K^0D^{(*)-}$  \\

$\overline B^0\to T_{cs\bar u\bar d}\overline D^{(*)0}$ & $T_3$ $V_{cb}V_{cs}^*$ & $\overline B^0\to D^{(*)+}K^- \overline D^{(*)0}$ & $\overline B^0\to D^{(*)0}\overline K^0\overline D^{(*)0}$   \\

$B^-\to T_{cs\bar u\bar u}\overline D^{(*)0}$ & $\sqrt{2} T_3$ $V_{cb}V_{cs}^*$ & $B^-\to D^{(*)0}K^-\overline D^{(*)0}$  \\

$\overline B^0\to T_{cs\bar d\bar d}D^{(*)-}$ & $\sqrt{2} T_3$ $V_{cb}V_{cs}^*$    & $\overline B^0\to D^{(*)+}\overline K^0D^{(*)-}$  \\
\hline
$\overline B_s^0\to T_{cd\bar u\bar s}\pi^0$ &${1\over\sqrt{2}}(T_3-T_2)$ $V_{cb}V_{ud}^*$ & $\overline B_s^0\to D_s^{(*)+}\pi^-\pi^0$ & $\overline B_s^0\to D^{(*)0}K^0\pi^0$  \\

$\overline B_s^0\to T_{cd\bar s\bar s}K^-$ & $\sqrt{2} T_3$ $V_{cb}V_{ud}^*$  & $\overline B_s^0\to D_s^{(*)+}K^0K^-$  \\

$B^-\to T_{cd\bar u\bar s}K^-$ &($T_1$ + $T_3$) $V_{cb}V_{ud}^*$ & $B^-\to D_s^{(*)+}\pi^-K^-$ & $B^-\to D^{(*)0}K^0K^-$  \\

$B^-\to T_{cd\bar u\bar u}\pi^0$ & ($T_1$ + $T_3$ - $T_2$) $V_{cb}V_{ud}^*$   & $B^-\to D^{(*)0}\pi^-\pi^0$  \\

$\overline B_s^0\to T_{cd\bar u\bar u}K^+$ & $\sqrt{2} T_1$ $V_{cb}V_{ud}^*$   & $\overline B_s^0\to D^{(*)0}\pi^-K^+$  \\

$\overline B_s^0\to T_{cd\bar u\bar s}\phi $ &$T_1$ $V_{cb}V_{ud}^*$  & $\overline B_s^0\to D_s^{(*)+}\pi^-\phi $ & $\overline B_s^0\to D^{(*)0}K^0\phi $  \\

$\overline B^0\to T_{cd\bar u\bar s}\overline K^{(*)0}$ & ($E$ + $T_1$) $V_{cb}V_{ud}^*$ & $\overline B^0\to D_s^{(*)+}\pi^- \overline K^{(*)0}$ & $\overline B^0\to D^{(*)0}K^0\overline K^{(*)0}$  \\

$\overline B^0\to T_{cd\bar u\bar u}\pi^+$ &  $\sqrt{2}$($E$ + $T_1$) $V_{cb}V_{ud}^*$ & $\overline B^0\to D^{(*)0}\pi^- \pi^+$  \\

$B^-\to T_{cs\bar u\bar u}K^{(*)0}$ & $\sqrt{2}T_2$ $V_{cb}V_{ud}^*$ & $B^-  \to D^{(*)0}K^-K^{(*)0}$  \\

$\overline B^0\to T_{cs\bar u\bar d}K^{(*)0}$ &  ($E$ + $T_2$) $V_{cb}V_{ud}^*$ & $\overline B^0\to D^{(*)+}K^- K^{(*)0}$ &  $\overline B^0\to D^{(*)0}\overline K^0K^{(*)0}$  \\

$\overline B^0\to T_{cs\bar u\bar u}K^+$ & $\sqrt{2}$ $E$ $V_{cb}V_{ud}^*$ & $  \overline B^0\to D^{(*)0}K^-K^+$  \\
\hline\hline
\end{tabular} \end{table}

It is observed from Table~\ref{tb:openc} that the $b\to c\bar cs$ channels can be classified into two categories according to their 
topological amplitudes, the $C$ one and the $T_3$ one. The decay channels sharing the same topological amplitude should have similar 
decay widths. Some of them are also related by the flavor symmetries, including 
\begin{align}
\text{isospin symmetry: } \ &A(B^-\to T_{\bar c\bar u ds}D^{(*)+}) = A(\overline B^0\to T_{\bar c\bar d u s}D^{(*)0}) \; , \\
&A(B^-\to T_{\bar c\bar u s s}D_s^{(*)+}) = A(\overline B^0\to T_{\bar c\bar d s s}D_s^{(*)+})  \; ,\\
&A(B^-\to T_{cs\bar u\bar d}D^{(*)-}) = A(\overline B^0\to T_{cs\bar u\bar d}\overline D^{(*)0}) \; , \\
&A(B^-\to T_{cs\bar u\bar u}\overline D^{(*)0}) = A(\overline B^0\to T_{cs\bar d\bar d}D^{(*)-}) \; .
\end{align}
These relations can be tested at future experiments. For the $b\to c\bar{c}s$ processes,  there are no processes related by the $U$-spin or $V$-spin symmetry as in the hidden-charm case. There are no $b\to c\bar{u}d$ processes related by the flavor SU(3) symmetry.

Three open-charm tetraquarks have been observed experimentally, including $T_{cs\bar{u}\bar d}$ observed in the channel $B^-\to T_{cs\bar u\bar d}D^{(*)-}$~\cite{LHCb:2020pxc}, $T_{\bar c\bar u ds}$ observed in $B^-\to T_{\bar c\bar u ds}D^{(*)+}$~\cite{LHCb:2022xob,LHCb:2022bkt}, and $T_{\bar c\bar d u s}$ observed in $\overline B^0\to T_{\bar c\bar d u s}D^{(*)0}$~\cite{LHCb:2022xob,LHCb:2022bkt}, respectively. It indicates that the amplitudes $C$ and $T_3$ should be large and the channels with amplitudes containing them should have a large chance to be observed in the near future. Therefore, we suggest the experimental searches for the following channels with priority, 
\begin{align}
&B^-\to T_{\bar c\bar u s s}D_s^{+} \to D_s^{-}K^-D_s^{+} \; , \\
&\overline B^0\to T_{\bar c\bar d s s}D_s^{+} \to D_s^{-}\overline K^0 D_s^{+} \; ,\\
&B^-\to T_{cs\bar u\bar u}\overline D^{0} \to D^{0}K^-\overline D^{0}  \; ,\\
&\overline B^0\to T_{cs\bar d\bar d}D^{(*)-} \to D^{(*)+}\overline K^0D^{(*)-} \; ,\\
&\overline B_s^0\to T_{cd\bar u\bar s}\pi^0 \to  D^{0}K^0\pi^0 (D_s^{+}\pi^-\pi^0 ) \; ,\\
&\overline B_s^0\to T_{cd\bar s\bar s}K^- \to D_s^{+}K^0K^-  \; ,\\
&B^-\to T_{cd\bar u\bar s}K^-  \to D^{0}K^0K^- ( D_s^{+}\pi^-K^- ) \; ,\\
&B^-\to T_{cd\bar u\bar u}\pi^0 \to D^{0}\pi^-\pi^0 \; .
\end{align}

\section{Conclusion}\label{sec:con}

We have analyzed the $B$ decay processes into a meson and a tetraquark, which can be hidden-charm or open-charm, by evaluating their topological amplitudes. It is found that some of the channels share the same topological amplitudes, which indicates that they have branching ratios of similar size and this can be examined experimentally. Among the hidden-charm tetraquark production processes, $\overline B^0\to T_{c\bar cu\bar d}K^- \to J/\psi \pi^+K^-$~\cite{Belle:2014nuw} and ${B^-\to T_{c\bar cs\bar u}\phi } \to J/\psi K^-\phi$ have been observed experimentally, so their amplitude $T_2$ should be large. Therefore, we suggest $\overline B^0\to T_{c\bar cs\bar d}\phi \to J/\psi \overline K^{(*)0}\phi$ and $B^-\to T_{c\bar cd\bar u}\overline K^{(*)0} \to  J/\psi \pi^-\overline K^{(*)0}$ should be searched experimentally with priority. Among the open-charm tetraquark production processes, $B^-\to T_{cs\bar u\bar d}D^{(*)-}$, $B^-\to T_{\bar c\bar u ds}D^{(*)+}$ and $\overline B^0\to T_{\bar c\bar d u s}D^{(*)0}$ have been observed experimentally, indicating that $T_3$ and $C$ should be large. Therefore, we suggest the priority channels $B^-\to T_{\bar c\bar u s s}D_s^{+} \to D_s^{-}K^-D_s^{+}$, $\overline B^0\to T_{\bar c\bar d s s}D_s^{+} \to D_s^{-}\overline K^0 D_s^{+}$, $B^-\to T_{cs\bar u\bar u}\overline D^{0} \to D^{0}K^-\overline D^{0}$
and $\overline B^0\to T_{cs\bar d\bar d}D^{(*)-} \to D^{(*)+}\overline K^0D^{(*)-}$ to be searched. 

\section*{Acknowledgement}

This work is supported in part by the National
Natural Science Foundation of China under Grant
Nos. 12005068, 11975112, and National Key Research
and Development Program of China under Contract No.
2020YFA0406400.







\end{document}